\documentclass[twocolumn]{aastex631}
\usepackage{savesym}
\savesymbol{tablenum}
\usepackage[print-unity-mantissa=false, separate-uncertainty, multi-part-units=single]{siunitx}
\DeclareSIUnit{\au}{AU}
\DeclareSIUnit\year{yr}

\restoresymbol{SIX}{tablenum}

\usepackage{amsmath}
\usepackage{amsfonts}
\usepackage{subfigure}
\usepackage{amssymb}
\usepackage{soul}

\begin{document}

\title{Atmospheric Mass Loss from TOI-1259~A~b, a Gas Giant Planet With a White Dwarf Companion}

\correspondingauthor{Morgan Saidel}
\email{msaidel@caltech.edu}

\author[0000-0001-9518-9691]{Morgan Saidel}
\affiliation{Division of Geological and Planetary Sciences, California Institute of Technology, Pasadena, CA 91125, USA}

\author[0000-0003-2527-1475]{Shreyas Vissapragada}
\affiliation{Carnegie Science Observatories, 813 Santa Barbara Street, Pasadena, CA 91101, USA}

\author[0000-0002-5547-3775]{Jessica Spake}
\affiliation{Carnegie Science Observatories, 813 Santa Barbara Street, Pasadena, CA 91101, USA}

\author[0000-0002-5375-4725]{Heather A. Knutson}
\affiliation{Division of Geological and Planetary Sciences, California Institute of Technology, Pasadena, CA 91125, USA}

\author[0000-0001-6025-6663]{Dion Linssen}
\affiliation{Anton Pannekoek Institute of Astronomy, University of Amsterdam, Science Park 904, 1098 XH Amsterdam, The Netherlands}

\author[0000-0002-0659-1783]{Michael Zhang}
\affiliation{Department of Astronomy \& Astrophysics, University of Chicago, Chicago, IL 60637, USA}

\author[0000-0002-0371-1647]{Michael Greklek-McKeon}
\affiliation{Division of Geological and Planetary Sciences, California Institute of Technology, Pasadena, CA 91125, USA}

\author[0000-0001-7144-589X]{Jorge P\'erez-Gonz\'alez}
\affiliation{Department of Physics and Astronomy, University College London,Gower Street, WC1E 6BT London, UK }

\author[0000-0002-9584-6476]{Antonija Oklopčić}
\affiliation{Anton Pannekoek Institute of Astronomy, University of Amsterdam, Science Park 904, 1098 XH Amsterdam, The Netherlands}

\begin{abstract}

The lack of close-in Neptune-mass exoplanets evident in transit surveys has largely been attributed either to photoevaporative mass loss or high-eccentricity migration. To distinguish between these two possibilities, we investigate the origins of TOI-1259 A b, a Saturn-mass (0.4 M$_J$, 1.0 R$_J$) exoplanet lying along the upper edge of the Neptune desert. TOI-1259 A b's close-in ($P$ = 3.48 days) orbit and low bulk density make the planet particularly vulnerable to photoevaporation. We studied the upper atmosphere of TOI-1259 A b using metastable helium~1083~nm transits observed with Palomar/WIRC and Keck/NIRSPEC. We report a band-integrated excess absorption of $0.395\pm{0.072}\%$ with Palomar/WIRC and a spectroscopically-resolved 
\SI{5.5(0.94)}{\kilo\meter\per\second} 
blueshifted absorption of $2.4\pm0.52\%$ ($T_1-T_4$) and $3.5\pm0.72\%$ ($T_2-T_3$) with Keck/NIRSPEC.
These measurements indicate the presence of an extended escaping atmosphere. Fitting these signals with a Parker wind model, we determine a corresponding atmospheric mass loss rate of log($\dot{M}$) = $10.2-10.65$ g/s for thermosphere temperatures between  $7900-8600$~K. This relatively low rate suggests that this planet would not have been significantly altered by mass loss even if it formed in-situ. However, the presence of a white dwarf companion, TOI-1259 B, hints that this planet may not have formed close-in, but rather migrated inward relatively late. Given the estimated parameters of the proto-white dwarf companion, we find that high-eccentricity migration is possible for the system.

\end{abstract}

\section{Introduction} \label{sec:intro}

\begin{figure}[ht] \label{fig:NeptuneDesert}
    \hspace*{-0.52cm}
    \centering
    \includegraphics[width=0.55\textwidth]{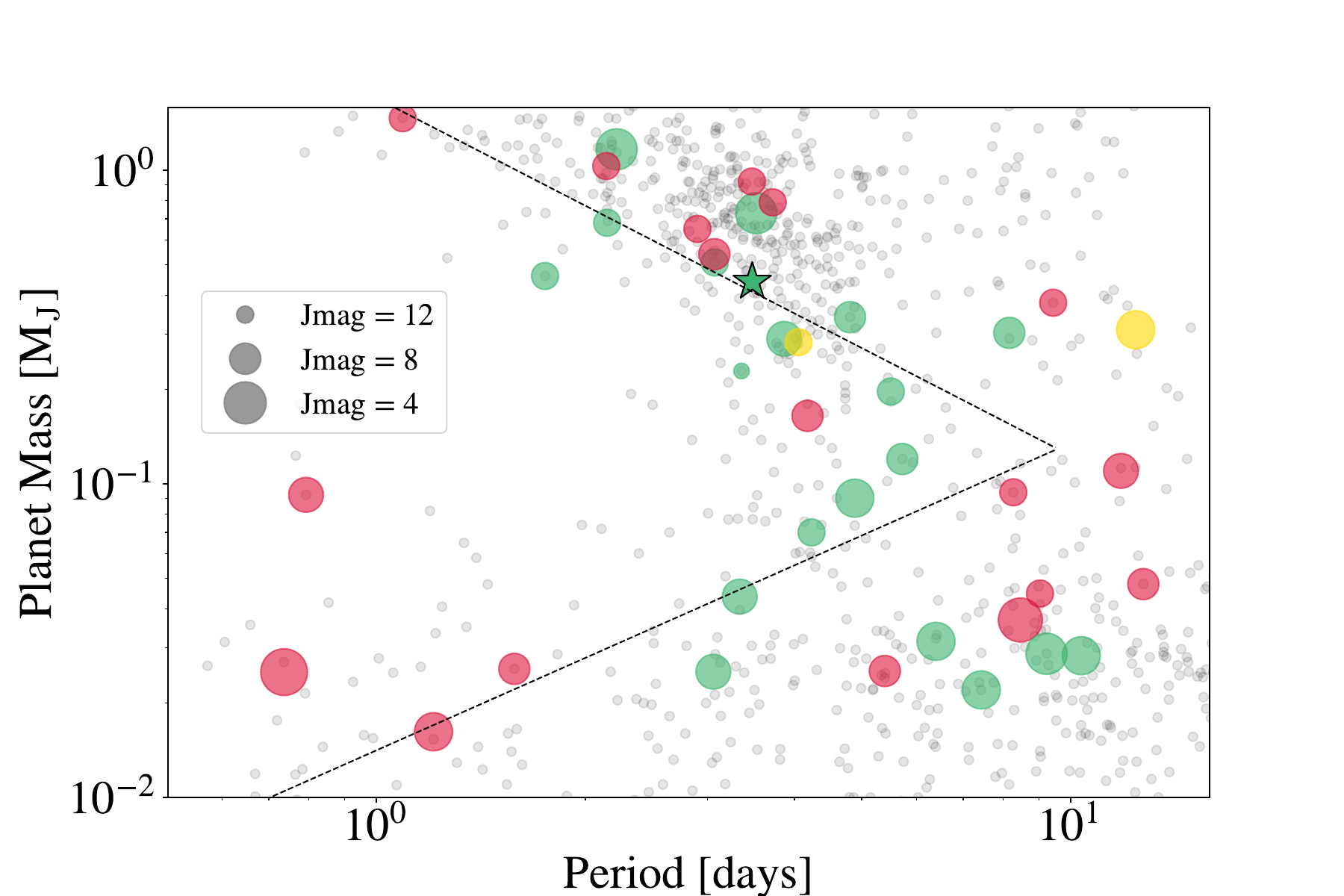}
    \caption{Mass versus period distribution for the sample of confirmed exoplanets as drawn from the NASA Exoplanet Archive on January 23, 2024 \citep{AkesonChen2013, ps}. The boundaries of the Neptune desert from \cite{MazehHolczer2016} are overplotted as black dotted lines. Green points indicate planets with published detections of atmospheric escape using the helium line, while red points indicate published non-detections, and yellow points denote tentative detections. The location of TOI-1259~A~b is indicated by a green star. Each point's size is scaled based on the host star's J magnitude (except for TOI-1259 which has a J magntiude of 10.2). For a full list of planets see \citet{GuilluyBourrier2023} and the review by \citet{DosSantos2023}; we also include detections for TOI-1268~b \citep{PerezGonzalezGreklek-McKeon2023}, HAT-P-67~ \citep{Gully-SantiagoMorley2023}, TOI-2134~b \citep{ZhangDai2023}, and the non-detection for LTT 9779~b \citep{EdwardsChangeat2023, VissapragadaMcCreery2024}.}
    
\end{figure}

Sub-Jovian exoplanets rarely reside on close-in orbits ($P$ $\leq$ 5 days). This gap in the exoplanet population is often referred to as the ``Neptune desert" (see Figure~\ref{fig:NeptuneDesert}), and its origin is thought to be linked to planet formation, migration and/or mass loss processes \citep{SzaboKiss2011, BeaugeNesvorny2013, LundkvistKjeldsen2016}. If the planets formed \textit{in situ} \citep{BatyginBodenheimer2016, BaileyBatygin2018} or migrated to their current positions early on via interactions with the gas disk \citep{IdaLin2008}, their atmospheres would have been exposed to significant quantities of high-energy radiation. High-energy radiation can drive strong atmospheric outflows, potentially stripping away the atmospheres of the least massive and most highly irradiated gas giants \citep{KurokawaNakamoto2014, LundkvistKjeldsen2016, ThorngrenLee2023}. This mechanism is likely responsible for sculpting the population of planets in the lower part of the Neptune desert \citep{OwenLai2018}. However, theoretical \citep{OwenLai2018, IonovPavlyuchenkov2018} and observational \citep{VissapragadaKnutson2022, GuilluyBourrier2023} studies suggest that planets along the upper edge of the desert are too massive for substantial photoevaporation of their envelopes. This suggests that the upper edge of the desert is likely primordial in nature. 

In the absence of significant atmospheric mass loss, it has been suggested that high-eccentricity migration (HEM) might produce the upper edge of the desert  \citep[e.g.][]{MatsakosKonigl2016, OwenLai2018}.  
This model proposes that after dissipation of the gas disk, dynamical interactions between the hot Jupiter progenitor (assumed to have formed far from the star) and other bodies in the system could significantly increase the planet's orbital eccentricity. If the orbit is eccentric enough, tidal interactions between the planet and the star will circularize the planet onto a new, close-in orbit \citep{RasioFord1996, OwenLai2018}. The upper edge of the desert is well matched by HEM models: less massive planets are more susceptible to tidal disruption during close pericenter passages, whereas more massive planets can survive these passages. This results in a mass-dependent minimum orbital separation for the final tidally circularized planet population. Lending additional credence to this idea, many close-in gas giants near the desert boundary have observed high obliquities \citep{BourrierAttia2023}; spin-orbit misalignment is thought to be a natural consequence of HEM \citep{NaozFarr2012, AlbrechtWinn2012, NelsonFord2017}. However, it is currently unclear what fraction of the planets along the upper edge of the desert underwent HEM \citep{FortneyDawson2021, JacksonDawson2023, YeeWinn2023}.

If HEM is common, this would have important implications for the mass loss histories of these close-in gas giants. Mass loss rates are predicted to be highest at early times, when stars are more active and planets may be inflated (from the heat of formation) and thus more vulnerable to photoevaporative mass loss \citep{JacksonDavis2012, Owen2019}. We might therefore expect planets that migrated late to be able to retain more of their atmosphere than those located on close-in orbits at earlier times.  

The recently-discovered planet TOI-1259 A b is an invaluable test case for distinguishing between HEM and photoevaporation scenarios. This gas giant has a mass of 0.44 M$_{J}$, a radius of 1.02 R$_{J}$, and is found on a relatively close-in ($P$ = 3.48 days) orbit around a K dwarf \citep{MartinEl-Badry2021}, placing it near the upper edge of the Neptune desert (Figure~\ref{fig:NeptuneDesert}). The planet's low density and short orbit render it an excellent candidate for observing photoevaporation in action. The system also includes a white dwarf companion (0.56 M$_{\odot}$, $a = 1648$~AU), which is estimated to have started with a mass of 1.59~M$_{\odot}$ and an orbital semi-major axis of $a\sim~900$~AU (assuming adiabatic mass loss) when it was on the main sequence \citep{MartinEl-Badry2021, FitzmauriceMartin2023}. The presence of the white dwarf allows for the system to have an unusually well-constrained age (4.8$_{-0.8}^{+0.7}$ \si{\giga\year}, which aligns well with gyrochronology estimates), as fitting the spectral energy distribution of the white dwarf provides a robust estimate of the white dwarf cooling age. However, perhaps most enticing about the white dwarf is that when it was on the main-sequence it may have been large enough and close enough to have excited TOI-1259 A b's orbital eccentricity (via e.g. the Kozai-Lidov mechanism) and caused it to migrate inward from a more distant formation location. As the planet migrated onto a close-in orbit, the resulting tidal circularization and reduction in semi-major axis would have terminated secular effects \citep{MartinEl-Badry2021}. 

%The evolution of the main sequence companion into a white dwarf would have then terminated the secular effects and, combined with tidal circularization, would have led to the planet's close-in circular orbit \citep{MartinEl-Badry2021}. 

In this study, we use observations of metastable helium (He$^*$) absorption at \SI{1083.3}{\nano\meter} to characterize the atmospheric mass loss rate of TOI-1259 A b and constrain its corresponding mass loss history. Planets with atmospheric outflows will have extended atmospheres, and may appear larger in transit when measured in this line \citep{OklopcicHirata2018, NortmannPalle2018, SpakeSing2018}. With over 15 detections of atmospheric mass loss to date (see Fig.~\ref{fig:NeptuneDesert}), this technique is currently the most widely successful method for measuring mass loss rates for transiting planets. In Section~\ref{sec:observ}, we describe our observations of the helium light curve of TOI-1259~A~b using the Wide-Field Infrared Camera (WIRC) at Palomar Observatory \citep{WilsonEikenberry2003} and the Near-Infrared Spectrograph (NIRSPEC) at Keck Observatory. In Section~\ref{sec:masslossmodeling}, we use our measurements of the excess He$^*$ absorption to estimate an atmospheric mass loss rate \citep{OklopcicHirata2018, DosSantosVidotto2022, LinssenOklopcic2022}. Lastly, we discuss the implication of this system's mass loss rate on it's atmospheric lifetime and the feasibility of HEM for the system in Section~\ref{sec:discussion}.

\section{Observations and Data Reduction} \label{sec:observ}

\subsection{Palomar/WIRC} \label{sec:paloobs}

We observed two transits of TOI-1259 A b on UT April 15 2022 and UT June 20 2022 with Palomar/WIRC. Although we attempted to observe a third transit on UT July 18 2022, weather conditions were poor and we were unable to collect usable data during the transit event. All transits were observed in an ultra-narrow band helium filter centered on the helium \SI{1083.3}{\nano\meter} line with a full-width at half-maximum of \SI{0.635}{\nano\meter} as discussed in \cite{VissapragadaKnutson2020}. We followed a similar procedure to the one outlined in our previous studies, including \cite{VissapragadaKnutson2020}, \cite{ParagasVissapragada2021}, and \cite{VissapragadaKnutson2022}. We first observed a helium arc lamp with the helium filter, and used this observation to place our target on the region of the detector where the position-dependent transmission function of the filter was centered on the helium line. We also used a custom beam-shaping diffuser for these observations, which produces a 3$\arcsec$ FWHM top hat point-spread function (PSF) for the target and any reference stars in the field of view, allowing us to improve our duty cycle by increasing integration time while minimizing time-correlated systematics \citep{StefanssonMahadevan2017, VissapragadaJontof-Hutter2020}. 
We performed standard astronomical image calibration steps including dark subtraction and flat-field correction. 
We then used dithered background frames to remove background OH emission features from our images. 

Once the image calibration was complete, we performed aperture photometry on the target and three comparison stars using the \texttt{photutils} package \citep{Bradley2023}. We used the same three comparison stars for both nights of data. To determine the optimal aperture size for each observation, we first used an average of the comparison star light curves to normalize our target star's light curve. We then used a moving median filter to remove $4\sigma$ outliers. We tested apertures from 5 to 20 pixels in radius in 1 pixel steps (pixel scale: 0\farcs25/px), and selected the aperture size that minimized the variance in the normalized and filtered target star data. We found that the optimal aperture radius was 9 pixels for both nights.

We model each transit light curve using the open-source \texttt{exoplanet} package \citep{Foreman-MackeyLuger2021zenodo, Foreman-MackeyLuger2021}. Following a procedure similar to the one outlined in \cite{VissapragadaKnutson2022}, we modeled the target with a limb-darkened light curve using the open-source \texttt{starry} package \citep{LugerAgol2019}. We then multiplied this model by a systematics model comprised of a linear trend in time and a linear combination of comparison star light curves, where the linear weights of the comparison stars were a free parameter with a uniform prior of $\mathcal{U}(-2,2)$ that was optimized as part of the fit. The wide uniform prior allows comparison stars to be assigned negative weights. This ensures that stars with weights consistent with zero are not biased towards positive values, which would artificially increase their contribution to the overall systematics model. We note that while none of the comparison stars in our model were assigned negative weights, one comparison star had a weight that was $1\sigma$ consistent with zero. Running our models again without this comparison star produces less than $1\sigma$ differences in our retrieved posteriors. 

In our systematics model, we initially included three additional parameters: the time-varying telluric water absorption proxy, the distance from the median centroid, and the airmass. Following the procedure outlined in \cite{VissapragadaKnutson2022}, we carried out fits using all possible combinations of these three covariates (including one without any additional covariates) and calculated the Bayesian Information Criterion (BIC) for each model \citep{Schwarz1978}. For the first night, the systematics model that minimized the BIC included the water absorption proxy parameter, and for the second night the model that minimized the BIC included the water absorption proxy and airmass parameters. We therefore adopted these models in the final joint model fit presented here.

The transit light curves in our fits are parameterized using the orbital period $P$, the ratio of the planetary to stellar radius $R_{p}/R_\star$, the impact parameter $b$, the epoch $T_{0}$, the ratio of the semi-major axis to stellar radius  $a/R_\star$, the quadratic limb darkening coefficients $u_{1}$, $u_{2}$, and a jitter term log($\sigma_{extra}$) to quantify the discrepancy between the photon noise and the observed variance in the data. We use fixed limb darkening coefficients calculated using \texttt{ldtk} \citep{HusserWende-vonBerg2013, ParviainenAigrain2015} and assuming a stellar effective temperature $T_\mathrm{eff}=$ \SI{4775}{\kelvin}, a surface gravity $\log(g)=$\SI{4.5}{\centi\meter\per\second\squared} and a metallicity [Fe/H] = -0.5 dex \citep{MartinEl-Badry2021}. We note that while we did also run our fits with free limb darkening coefficients, our retrieved parameters did not change appreciably, so we therefore adopt the fixed limb darkening coefficients for the final version of our fit. We used the \texttt{pymc3} \citep{SalvatierWieckia2016} No U-Turn Sampler \citep[NUTS; ][]{HoffmanGelman2011} to sample the posterior probability distribution. For each of the fits to the individual nights of WIRC data, we ran four chains with 1,500 tuning steps and 2,500 draws per chain. We list the prior and posterior ranges for the full set of astrophysical model parameters for these fits in Table~\ref{tab:priors}.

\begin{figure*}[ht] 
    \centering
    \includegraphics[width=0.9\textwidth]{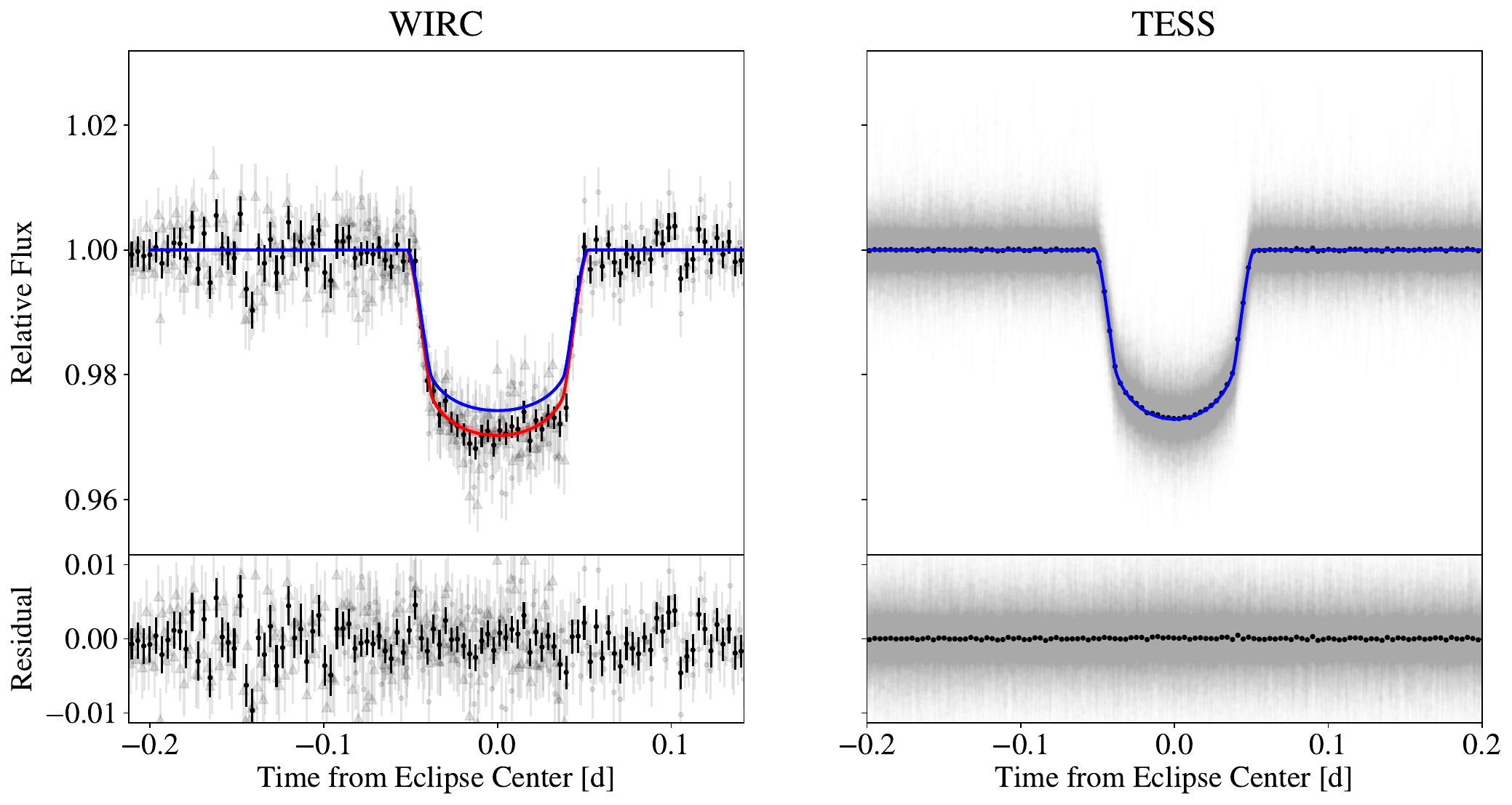}
    \caption{Transit light curve (top left) and residuals (bottom left) for both WIRC transit observations of TOI-1259~A~b. Unbinned data from the first and second nights of observation are shown as gray circles and triangles, respectively. Data from both nights are binned together in 10 minute intervals, which are shown as black circles. The red curve is the best-fit WIRC transit model from the joint fit, with red shading indicating the 1$\sigma$ uncertainty in the transit depth. The blue curve is the same transit model with $(R_p/R_{star})^2$ fixed to the value obtained by fitting the broadband \textit{TESS} data. The corresponding \emph{TESS} transit light curve (top right) and residuals (bottom right) are shown for comparison. Unbinned data are plotted in gray, with data points binned to 10 minutes shown in black. The best-fit \textit{TESS} transit model is overplotted in blue.  }
    \label{fig:joint fit}
\end{figure*}

After fitting the individual transit light curves fits for each night, we find a transit depth of $2.888_{-0.093}^{+0.090}\%$ for the first night and $2.75_{-0.14}^{+0.14}\%$ for the second night. These depths are consistent within 1$\sigma$, and we therefore conclude that the helium signal did not vary significantly over the two months that separated our two observations.  
We then jointly fit both nights of data with a common transit depth. We also simultaneously fit a light curve to the \textit{TESS} data \citep[two minute cadence, downloaded using the \texttt{lightkurve} package;][]{LightkurveCollaborationCardoso2018} from multiple sectors over a two year period to further constrain the transit shape and provide a reference transit depth that we can use to search for excess absorption in the helium bandpass. This methodology was previously used to detect an excess absorption signal in the He$^*$ line for HAT-P-18 b \citep{ParagasVissapragada2021}, which was later confirmed by a \emph{JWST} observation spanning the 1083 nm helium line \citep{FuEspinoza2022, Fournier-TondreauMacDonald2024}. We followed the same procedure as described for the single night fits, except we use free limb darkening coefficients in the \textit{TESS} bandpass. In this joint fit, we ran four chains with 5,000 tuning steps per chain and 5,000 draws. We found that our joint fit with uniform limb darkening coefficients results in retrieved coefficients that are within 1$\sigma$ of our calculated limb darkening coefficients, and similar to the single night fits, we proceeded with the calculated, fixed limb darkening coefficients in our final model. For all fits we found that the Gelman-Rubin parameter $\hat{R} < 1.01$ for all sampled parameters, which indicates convergence \citep{GelmanRubin1992}.  
The priors and posteriors for this fit are listed in Table~\ref{tab:priors} and the corresponding transit light curves are shown in Figure~\ref{fig:joint fit}.

\begin{table*}[!ht]
    \centering
    
    \tablewidth{300pt} 
    \caption{Priors and posteriors for the joint fit of both Palomar/WIRC transits and the \textit{TESS} data.} 
    \hspace*{-2.0cm}
    \begin{tabular}{ccccc}
    \hline
    \hline
      Parameter & Priors & Posterior & Posterior &  Posterior  \\
      &  & (night 1) & (night 2) & (joint) \\
      \hline
      \hline
         $P$ & $\mathcal{N}(3.477978, 0.0000019)$ & $3.4779781_{-0.0000014}^{+0.0000014}$ & $3.4779794_{-0.0000015}^{+0.0000015}$ & $3.47797926_{-0.00000017}^{+0.00000017}$\\
         
         ${R_p/R_*}_{WIRC}$ & $\mathcal{U}(0, 0.25)$ & $0.1579_{-0.0026}^{+0.0024}$ & $0.1542_{-0.0040}^{+0.0039}$ & $0.1602\pm{0.0019}$\\
         
         ${R_p/R_*}_{TESS}$ & $\mathcal{U}(0, 0.5)$ & - & - & $0.14910_{-0.00031}^{+0.00038}$\\
         $b$ & $\mathcal{N}(0.065, 0.055)$ & $0.062_{-0.053}^{+0.051}$ & $0.070_{-0.057}^{+0.053}$ & $0.082_{-0.053}^{+0.038}$\\
         $a/R_*$ & $\mathcal{N}(12.314, 0.056)$ & $12.313_{-0.051}^{+0.053}$ & $12.324_{-0.055}^{+0.055}$  & $12.300_{-0.035}^{+0.029}$\\
         ${u_1, u_2}_{TESS}$ & \cite{Kipping2013} & - & - &  $0.473_{-0.016}^{+0.017}$,~$0.16_{-0.04}^{+0.04}$\\
         log($\sigma_{extra}$) & $\mathcal{U}(10^{-6}, 10^{-2})$& $0.00299_{-0.00036}^{+0.00037}$& $0.00321_{-0.00037}^{+0.00038}$&  $0.00221_{-0.00043}^{+0.00041}$, $0.00250_{-0.00039}^{+0.00040}$\\
         $\delta_{mid}$ & - & $0.00365_{-0.00093}^{+0.00090}$ & $0.0023_{-0.0014}^{+0.0014}$ & $0.00395_{-0.00072}^{+0.00072}$ \\
         \hline
         \hline
    \end{tabular}
    \tablecomments{The fixed WIRC limb darkening coefficients are $u_1, u_2 = 0.35, 0.14$ calculated using \texttt{ldtk} \citep{HusserWende-vonBerg2013, ParviainenAigrain2015}. }
    \label{tab:priors}
\end{table*}

\subsection{Keck/NIRSPEC}

We observed a full transit of TOI-1259~A~b with Keck/NIRSPEC on UT June 27th, 2022 using the $Y$ band
filter in the high resolution mode. Observations were obtained using the \ang{;;0.288}$\times$ \ang{;;12} slit, which has a slit resolution of 37,500. We used \SI{300}{\second} exposures with an ABBA nodding pattern to subtract the background. To analyze and reduce the data, we use the pipeline described in \citet{ZhangKnutson2021,ZhangKnutson2022}, which we summarize here. 
We identify bad pixels as those that deviate by more than 5$\sigma$ from an image-wide mean of pixels in each dark. Bad pixels present in more than half of all individual dark frames are marked as bad in the master dark. We then produce a master flat, which we use to calibrate A-B subtracted images for each A/B pair. After zeroing out all pixels more than 10 pixels away from the trace (a region which notably includes the negative trace), we used optimal extraction to retrieve the 1D spectrum and associated errors. To produce a wavelength solution, we compute a template using model tellurics and a model stellar spectrum from the PHOENIX grid \citep{HusserWende-vonBerg2013}. We then shift the template stellar spectrum to account for Earth's velocity relative to TOI-1259~A on the night of observation and used the shifted spectrum to obtain wavelength solutions for each individual spectrum.

After obtaining the wavelength corrected 1D spectra, we used \texttt{molecfit} \citep{SmetteSana2015} to remove telluric absorption features and obtain the spectral resolution ($R\sim28000$). We then constructed a spectral grid, in which we interpolate all spectra onto a common wavelength grid. Using a notch filter, we corrected the spectral grid for detector fringing. We then constructed a residuals spectrum by taking the log of the spectral grid and subtracting off the mean of every row and column, resulting in every pixel displaying the fractional flux variation at that wavelength and timestamp. 

To track the excess absorption, we take every column (wavelength) of the residuals image and subtract the mean out-of-transit portion of the residuals, defined as the residuals preceding first contact $T_1$ and following fourth contact $T_4$ ($T_1-T_4$ is considered in-transit), resulting in an excess absorption at every wavelength. 
Given that the residuals image is in log space, subtracting the mean out-of-transit portion of the residuals is equivalent to dividing by the geometric mean. In order to remove continuum variations, we fit a third-order polynomial to a copy of 
the residuals spectrum  with all strong lines (including both stellar and telluric absorption lines) masked out. We fit the polynomial to each row with respect to the column number (wavelength) of the masked residuals spectrum. We then subtract the polynomial continuum fit from the original residuals spectrum. We find that in the normalized spectral timeseries there is still residual variability at the locations of telluric lines, and therefore mask them in the final residuals spectrum. We separately confirm that stellar lines, including the stellar helium line, are not variable and as a result we do not mask them in the final residuals spectrum.

\begin{figure}[ht]
    \centering
    \subfigure{\includegraphics[width=0.495\textwidth]{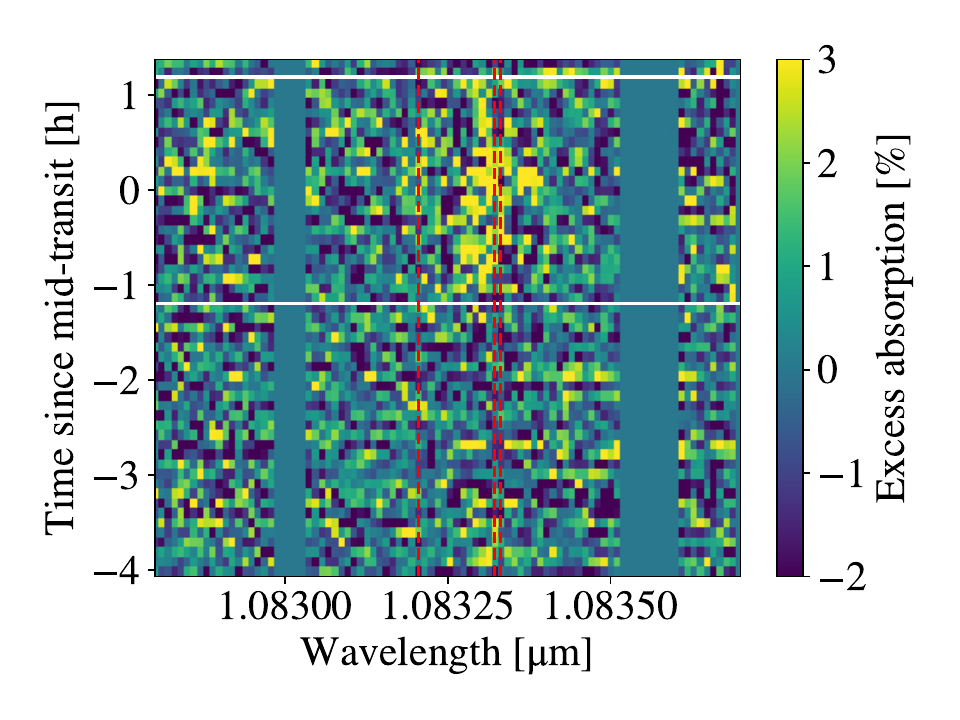}}\qquad
    \subfigure{\includegraphics[width=0.495\textwidth]{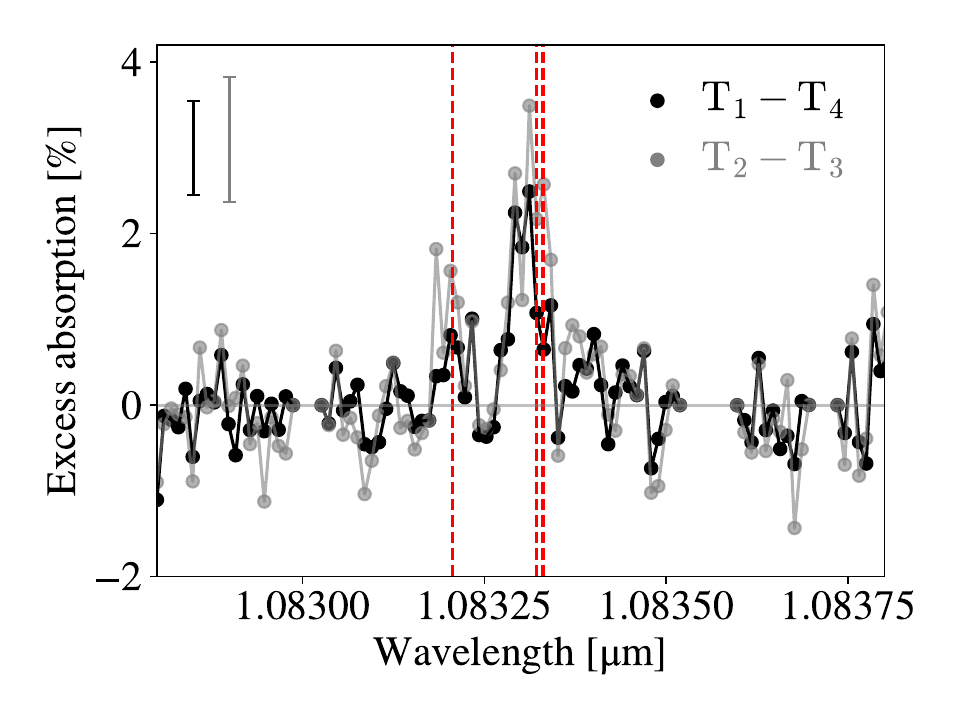}}
    \caption{Top: Excess absorption of TOI-1259~A~b, in percent, in the planetary rest frame as a function of time and wavelength, as observed with Keck/NIRSPEC. The white horizontal lines mark the beginning (bottom) and end (top) of transit. The red vertical dashed lines mark the locations of the three helium lines. We mask out strong telluric and stellar lines as part of our analysis, which are seen as gaps at 1.0830057 \si{\micro\meter} (stellar silicon line) and 1.08351 \si{\micro\meter} (water line).
    Bottom: Excess absorption spectra in the planetary rest frame as a function of wavelength. The black line represents the excess absorption spectrum calculated using all in-transit data ($T_1-T_4$). The gray line represents the average excess absorption spectrum calculated for the time period where the planet is fully in front of the star ($T_2-T_3$). Representative error bars for each spectrum are shown in the top left corner.}
    \label{fig:Keck_residuals}
\end{figure}

\begin{figure}[h!]
    \centering
    \includegraphics[width=0.495\textwidth]{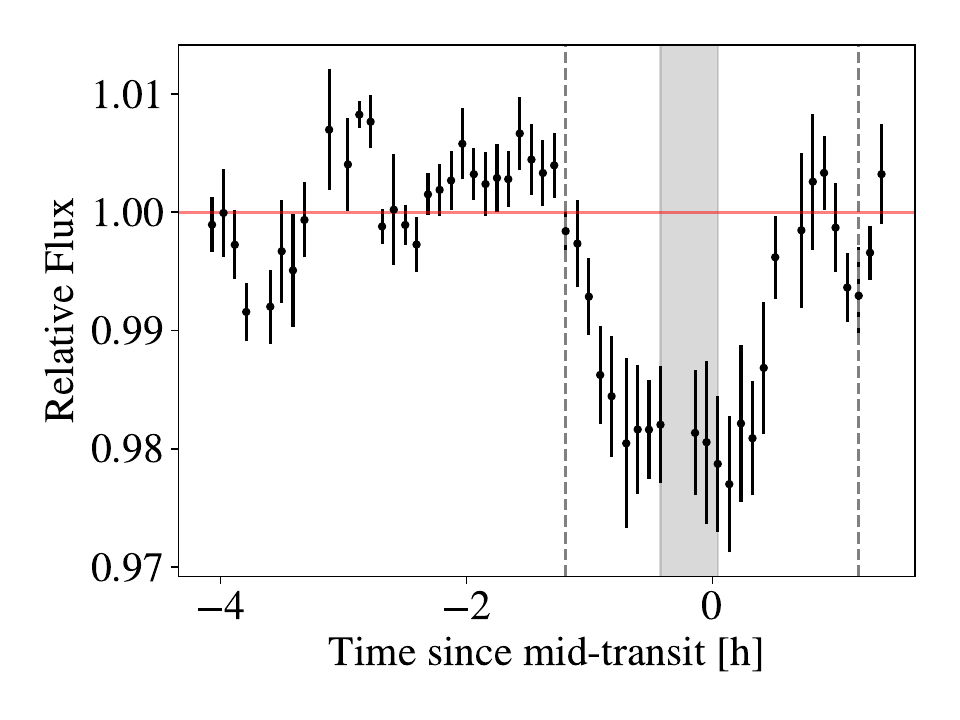}
    \caption{The band-integrated light curve of TOI-1259~A~b, integrated within \SI{0.75}{\angstrom} of the main peak in the helium triplet (10833.0 \si{\angstrom}). Gray dashed lines mark the beginning and end of transit. The gray shaded region in the band-integrated light curve denotes a focusing issue. }
    \label{fig:Keck_lightcurve}
\end{figure}

As a next step, we shifted the spectrum onto the planetary rest frame by computing the radial velocity of the planet relative to the star using each epoch's barycentric Julian date. We then linearly interpolated the spectrum onto a common wavelength grid. The final excess absorption spectrum as a function of time and wavelength is shown in the top panel of Figure~\ref{fig:Keck_residuals}, while the bottom panel shows the average in-transit excess absorption spectrum in the planet rest frame. Figure~\ref{fig:Keck_lightcurve} shows the band-integrated light curve in a \SI{0.75}{\angstrom} bandpass centered on the \SI{10833.0}{\angstrom} main peak of the helium line. We note that the pre-ingress and post-egress baseline do not deviate significantly from unity, consistent with our previous conclusion that we do not detect any statistically significant variations in the stellar helium line.

\begin{figure}[h!]
    \centering
    \includegraphics[width=0.495\textwidth]{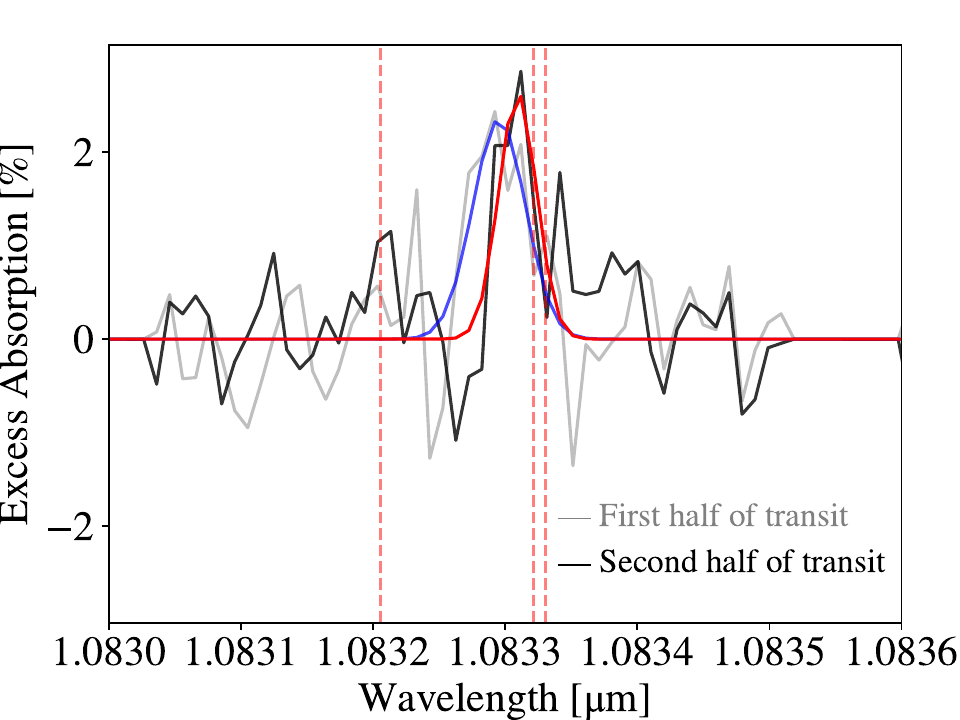}
    \caption{The average excess absorption during the first half of transit ($T_1$ to mid-transit, gray) and second half of transit (mid-transit to $T_4$, black). The best-fit Gaussian model for the first half of transit is shown in blue, the best-fit Gaussian model for the second half of transit is shown in red. Red vertical dashed lines denote the rest wavelengths of the three helium lines.}
    \label{fig:Keck_firsthalfsecondhalf}
\end{figure}

\section{Results \& mass loss Modeling} \label{sec:masslossmodeling}

We measure a transit depth $(R_p/R_{star})^2$ in the helium line of $0.02969_{-0.00071}^{+0.00072}$ with Palomar/WIRC and $0.025746_{-0.000093}^{+0.000102}$ in the \textit{TESS} light-curve fit. This corresponds to a mid-transit excess absorption of $0.395\pm{0.072}\%$, indicating the presence of an extended escaping atmosphere on TOI-1259 A b. With Keck/NIRSPEC we measure a wavelength-resolved excess absorption signal with a peak value of $2.4\pm0.52\%$ in the line core, which appears to be blueshifted.  We quantify the magnitude of the blueshift by fitting a Gaussian function to the Keck excess absorption spectrum using \texttt{lmfit} \citep{NewvilleStensitzki2016} and find that the helium line is blueshifted by \SI{5.5(0.94)}{\kilo\meter\per\second}. In order to facilitate comparisons with the Palomar/WIRC mid-transit excess absorption, we additionally report the excess absorption signal constructed only using data between second and third contact ($T_2-T_3$, see Figure~\ref{fig:Keck_residuals}). This excess absorption spectrum has a peak value of $3.5\pm0.72\%$, and is more reflective of the behavior at mid-transit. 

In the top panel of Fig.~\ref{fig:Keck_residuals}, there are some hints that the location of the excess absorption peak may be changing as the transit proceeds.
This kind of time-evolving structure is a common feature in many 3D models \citep[e.g.,][]{WangDai2021, NailOklopcic2023}. We evaluated the statistical significance of this apparent wavelength shift by 
dividing the 
transit in half and comparing the location of the averaged excess absorption peak during the first half ($T_1$ to mid-transit) versus the second half (mid-transit to $T_4$) of the transit, as shown in Figure~\ref{fig:Keck_firsthalfsecondhalf}. Using weighted least squares minimization with \texttt{lmfit} \citep{NewvilleStensitzki2016}, we fit a model Gaussian to determine the location of each excess absorption peak.  We find that the peak of the excess absorption occurs at  $10832.95\pm0.04$ \si{\angstrom} during the first half of the transit corresponding to a blueshift of {\SI{8.2(1.2)}{\kilo\meter\per\second}} from the main triplet helium peak, whereas the second half has a peak at 
$10833.09\pm0.03$ \si{\angstrom}, corresponding to a blueshift of {\SI{4.3(0.92)}{\kilo\meter\per\second}} from the main triplet helium peak. These two values differ by 2.8$\sigma$, indicating that the location of the peak excess absorption appears to shift over the course of the transit. We note that the width of these two lines ($0.19\pm0.04$~\si{\angstrom} for the first half of transit, $0.14\pm0.03$~\si{\angstrom} for the second half of transit) as well as the width of the helium line for the full transit ($0.18\pm0.03$~\si{\angstrom}) are consistent with instrumental broadening ($0.18$~\si{\angstrom}).
Although a more detailed 3D model analysis is beyond the scope of this paper, we provide the data behind Figures~\ref{fig:Keck_residuals}-\ref{fig:Keck_firsthalfsecondhalf} in order to facilitate future 3D studies.

\begin{figure}[h!]
    \centering
    \includegraphics[width=0.5\textwidth]{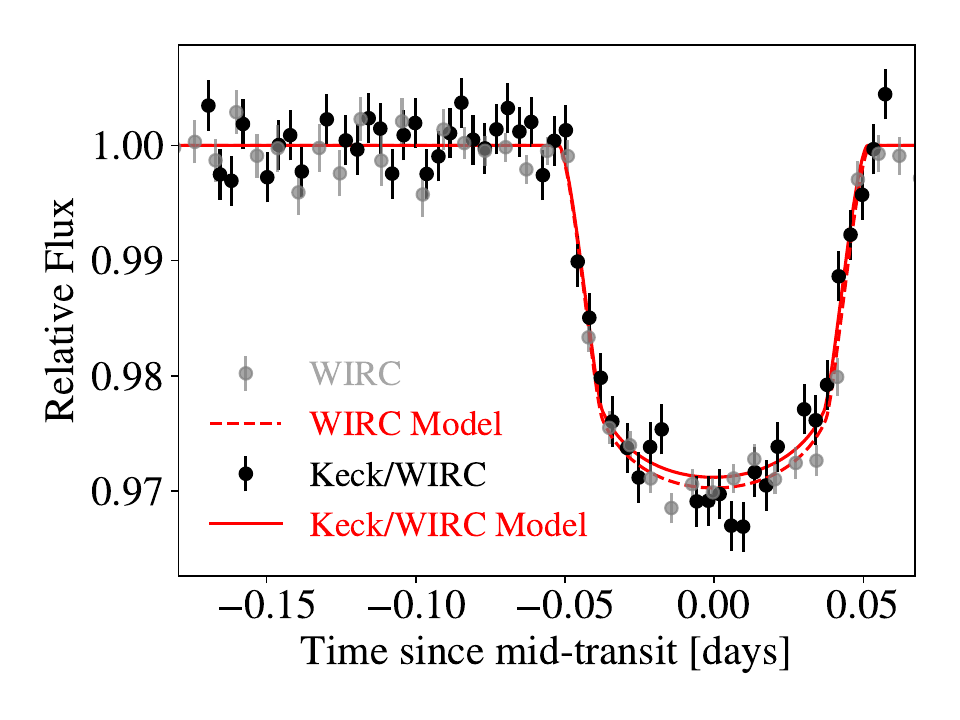}
    \caption{Synthetic Keck timeseries in the Palomar/WIRC helium filter (black points) as compared to our Palomar/WIRC transit observation (grey points).  We overplot the best-fit transit model for the Keck timeseries as a red solid line and the best-fit transit model for the Palomar/WIRC timeseries as a red dashed line.}
    \label{fig:Keck_convovledwithWIRC}
\end{figure}

Next, we check for time variability in TOI-1259~A~b's helium signal by comparing the Keck and Palomar excess absorption measurements. We convolved our Keck/NIRSPEC spectrum with the Palomar/WIRC helium filter by first trimming the Keck spectrum around the helium feature, which we identify using the \texttt{sunbather} best-fit helium models (see Section~\ref{sec:masslossmodeling}). We then pad zeros onto either end of the trimmed spectrum out to the full wavelength range of the Palomar/WIRC filter.  We then integrate the spectrum in the Palomar bandpass following the methodology outlined in \citet{VissapragadaKnutson2020}, and find an excess absorption of $0.185\pm0.041\%$ from the Keck $T_1-T_4$ excess absorption spectrum and a Palomar/WIRC-predicted excess absorption of $0.301\pm0.055\%$ from the Keck $T_2-T_3$ excess absorption spectrum. 
 Our $T_2-T_3$ excess absorption, which is more reflective of the behavior at mid-transit, is consistent with the Palomar/WIRC mid-transit excess absorption of $0.395\pm0.072\%$ at the 1$\sigma$ level. We also separately create a simulated timeseries for our Keck data in the WIRC helium bandpass by convolving each individual Keck spectrum with the WIRC transmission function and adding in a white light curve based on our \textit{TESS} model fits. We plot the resulting light curve in Figure~\ref{fig:Keck_convovledwithWIRC}, and compare it to our measured WIRC transit light curve. Following a similar methodology as described in Section~\ref{sec:paloobs}, we fit the simulated Keck transit lightcurve using \texttt{exoplanet} \citep{Foreman-MackeyLuger2021, Foreman-MackeyLuger2021zenodo} and \texttt{pymc3} \citep{SalvatierWieckia2016} and determine a mid-transit excess absorption of $0.300\pm0.071\%$.  
This best-fit value is nearly identical to our band-integrated $T_2-T_3$ excess absorption value, but has a modestly higher uncertainty.  This is unsurprising, as we estimate the uncertainty in this version of the fit using the scatter in the band-integrated timeseries prior to the start of ingress modified by an error scaling term that is optimized in the fit. We find that the best-fit uncertainty is $1.2$ times the scatter in the pre-ingress data. For the $T_2-T_3$ excess absorption spectrum we set the uncertainty in each wavelength bin equal to the standard deviation of the excess absorption spectrum at wavelengths outside the helium line. We conclude that there is no significant variability in the underlying helium signal between the Keck/NIRSPEC and Palomar/WIRC measurements.

Using the excess absorptions derived from both our Palomar fits and Keck measurements, we place a constraint on the mass loss rate by modeling the outflow using \texttt{sunbather}, following the methodology in \cite{LinssenOklopcic2022, LinssenShih2024}.
We begin by defining a grid in $\dot{M}$ and $T_0$ and generating a one-dimensional isothermal Parker wind model \citep{OklopcicHirata2018, LamponLopez-Puertas2020, DosSantosVidotto2022} for each grid point. We consider mass loss rates between $10^{9}$  and $10^{11}$~g~s$^{-1}$, and thermosphere temperatures between $5000-11000$~K. We assume a hydrogen fraction of 0.9 (i.e., 90\% hydrogen and 10\% helium by number) and calculate one-dimensional density and velocity profiles for each combination of $\dot{M}$ and $T_0$. We then use the open-source \texttt{Cloudy} package \citep{FerlandKorista1998, FerlandChatzikos2017} for each $T_0$ profile to calculate the corresponding heating and cooling rates as a function of altitude and ionization structure for each $T_0$. These heating and cooling rates are used to derive a new temperature structure, which we then use as the input for a new \texttt{Cloudy} simulation run. We repeat this process until the heating and cooling rates are balanced. We then calculate the mean and standard deviation of the temperature profile in the helium line-forming region. We reject models with a large difference between this mean temperature and $T_0$ as unphysical \citep{LinssenOklopcic2022}.

\begin{figure*}[]
    \centering
    \includegraphics[]{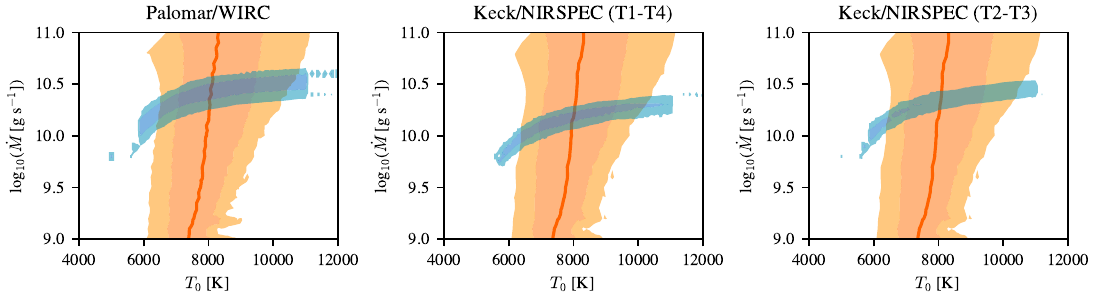}% <-- added
    \caption{The mass loss and temperature model results from \texttt{Cloudy} models for the measured Palomar/WIRC excess absorption (left) and Keck/NIRSPEC excess absorption ($T_1-T_4$: middle; $T_2-T_3$: right) using HD 40307. The orange line and orange shading represents the 2$\sigma$ (light orange) and 1$\sigma$ (dark orange) confidence intervals from the constrained temperature structure. The blue line and blue shading indicates the 2$\sigma$ (light blue) and 1$\sigma$ (dark blue) best-fit models of the mass loss.}      \label{fig:Cloudy_TOI1259}
\end{figure*}

For the remaining subset of radiatively self-consistent models, we use \texttt{Cloudy} to compute the density of metastable helium as a function of radial distance. Using this density profile, we then calculate the corresponding helium absorption signal during transit and integrate over the $0.635$~nm WIRC bandpass and separately convolve with the resolution of Keck NIRSPEC-Y at the metastable helium wavelengths to predict mass loss rates for the Palomar excess absorption and Keck excess absorption respectively. As shown in Figure~\ref{fig:Cloudy_TOI1259}, we find that our measured excess helium absorption with Palomar/WIRC is best-matched by models with a mass loss rate of $\log{\dot{M}} = 10.40^{+0.15}_{-0.20}$~g~s$^{-1}$ and a thermosphere temperature of $T_0 = 8400^{+1900}_{-1800}$~K. The measured $T_1-T_4$ excess helium absorption with Keck/NIRSPEC is best-matched by models with a mass loss rate of $\log{\dot{M}} = 10.20^{+0.10}_{-0.25} $~g~s$^{-1}$ and a thermosphere temperature of $T_0 = 8100^{+1900}_{-1600}$~K, while the measured $T_2-T_3$ excess helium absorption with Keck/NIRSPEC is best-matched by models with a mass loss rate of $\log{\dot{M}} = 10.30^{+0.15}_{-0.25}$~g~s$^{-1}$ and a thermosphere temperature of $T_0 = 7900.0^{+2100}_{-1700}$~K. Both Keck/NIRSPEC mass loss rates are within approximately 1$\sigma$ of the Palomar mass loss rate. We note that the $T_2-T_3$ Keck/NIRSPEC mass loss rate is  closer to the measured Palomar/WIRC mass loss rate than the $T_1-T_4$ value.  This is likely because the average excess absorption calculated when the planet is fully in front of the star ($T_2-T_3$) is a closer match to the value obtained by fitting the full light curve with a transit model.

\begin{figure}[h!]
    \centering
    \centering
    \subfigure{\includegraphics[width=0.49\textwidth]{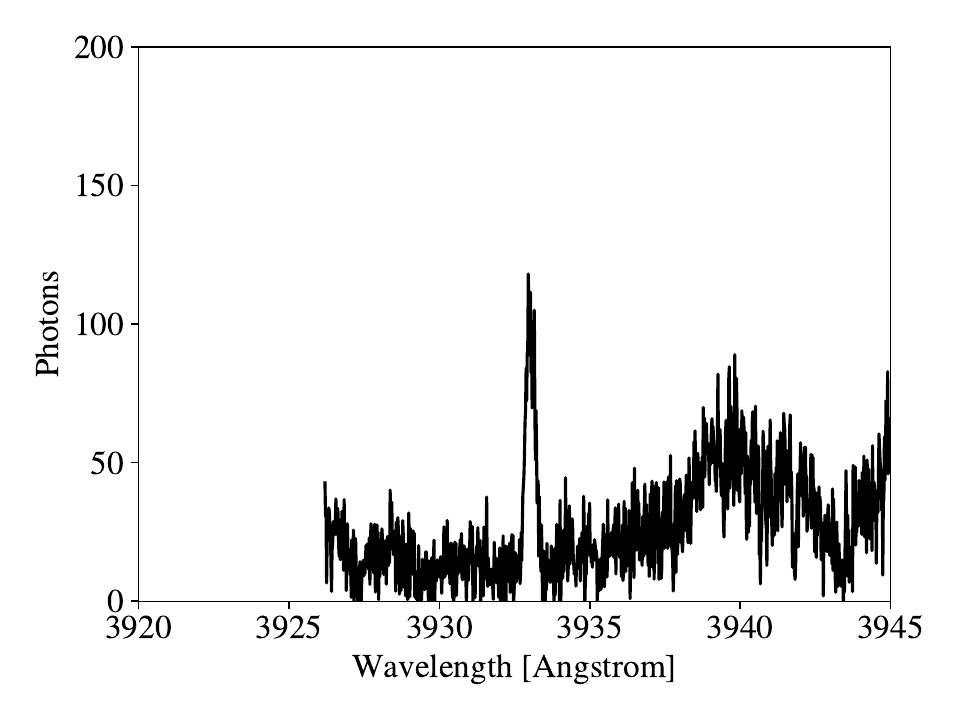}}\qquad
    \subfigure{\includegraphics[width=0.49\textwidth]{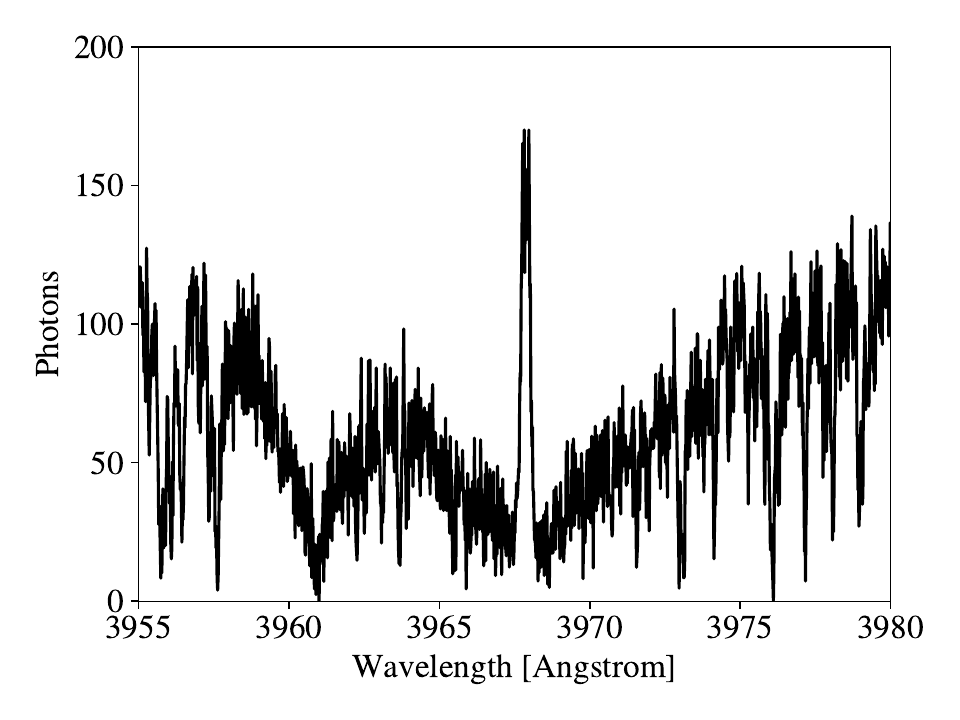}}
    \caption{Optical spectrum of TOI-1259~A obtained during a single epoch of observation with HIRES on UT November 28 2019. The Calcium II K (top) and H (bottom) lines are used to derive a log~$R'_{\mathrm{HK}}$ of -4.5 for the planet (Howard Isaacson, private communication) following the methodology in \citet{IsaacsonKane2024}. }
    \label{fig:TOI1259_CAHKplots}
\end{figure}

Given the unknown stellar spectrum of the host star, these mass loss rates are estimated using the proxy MUSCLES  \citep{YoungbloodFrance2017} stellar spectrum of HD 40307, which proves a close match in effective temperature, surface gravity, spectral type, metallicity and age to TOI-1259~A. However, an optical spectrum of TOI-1259~A obtained with the High Resolution Infrared Spectrograph (HIRES) on Keck (see Figure~\ref{fig:TOI1259_CAHKplots}) reveals a log~$R'_{\mathrm{HK}}$ of -4.5 (Howard Isaacson, private communication), indicating an active chromosphere. TOI-1259~A may be quite active for its age, and falls near the edge of the rotation-log~$R'_{\mathrm{HK}}$ for K stars \citep{Toledo}. Given the apparent enhanced stellar activity of the star, we repeat our mass loss models using the more active proxy stellar spectrum of $\epsilon$ Eridani \citep{YoungbloodFrance2017}. We find a maximum mass loss rate of $\log{\dot{M}} = 10.65^{+0.15}_{-0.30}$~g~s$^{-1}$ and a thermosphere temperature of $T_0 = 8600^{+2400}_{-2200}$~K. We note that the true mass loss rate is likely in between our mass loss rates derived using HD 40307 and $\epsilon$ Eridani, as the log~$R'_{\mathrm{HK}}$ measurement is based on a single epoch of HIRES observation and if the star has a strong activity cycle or rotational modulation, the HIRES observation could have occurred near a maximum of either scenario, driving the activity of the star to appear higher than it would with repeat observations. 

\section{Discussion}\label{sec:discussion}

\subsection{Day-Night Advection Shapes Blueshifted He$^*$ Absorption}

TOI-1259~A~b displays a net blueshifted excess absorption (see Fig.~\ref{fig:Keck_residuals}). Blueshifted excess helium absorption signals are often accompanied by post-egress absorption \citep{NortmannPalle2018, SpakeSing2018, AllartBourrier2019, KirkAlam2020}. This outflow morphology is predicted to occur in systems with strong stellar winds, which confine the planetary outflow into a comet-like blueshifted tail trailing the planet. However, the lack of post-egress absorption in our data disfavors this scenario as the origin of the net blueshift. Instead, it is likely that advection of gas from the hotter day side to the cooler night side is creating the blueshifted signal. The uppermost layers of the dayside atmosphere are heated by XUV radiation, creating a day-night temperature and pressure gradient that drive the outflowing gas towards the low pressure region on the night side of the planet. This results in a net blueshifted signal along the line of sight during transit \citep{WangDai2021,NailOklopcic2023}. 

\citet{WangDai2021} modeled this day-night advection and found that they were able to reproduce the blueshifted signal observed on WASP-69 without having to invoke a significant stellar wind induced tail. Although WASP-69 was subsequently shown to possess an extended tail \citep{TylerPetigura2024}, similar observations of HD 189733~b \citep{SalzCzesla2018, GuilluyAndretta2020, ZhangCauley2022}, HD 209458~b \citep{Alonso-FlorianoSnellen2019}, and HAT-P-11~b \citep{AllartBourrier2018} indicate that all three of these planets display blueshifted signals without significant post-egress absorption, and the authors of these studies invoke day-night side winds as the likely explanation.

\subsection{TOI-1259~A~b is not significantly affected by photoevaporation}

Using our retrieved mass loss rates, we estimate the planet's cumulative maximum mass loss by first calculating a mass loss efficiency $\epsilon$ (the efficiency with which the atmosphere converts incident radiation into heat). Following the procedure outlined in \cite{VissapragadaKnutson2022_maxmassloss}, and adopting our maximum mass loss rate using HD 40307 ($\log{\dot{M}} = 10.40^{+0.15}_{-0.20}$~g~s$^{-1}$) and $\epsilon$ Eridani ($\log{\dot{M}} = 10.65^{+0.15}_{-0.30}$~g~s$^{-1}$) we obtain a heating efficiency of 0.075$_{-0.031}^{+0.027}$ (HD 40307) and 0.054$_{-0.022}^{+0.049}$ ($\epsilon$ Eridani). Assuming this mass loss efficiency is constant, we calculate the maximum fraction of the planet mass that has been lost to photoevaporation following the approach detailed in \cite{VissapragadaKnutson2022_maxmassloss}. In order to calculate, the maximum fraction of the planet's mass lost, we adopt an integrated XUV flux of $E_{XUV}~\sim~10^{46}$~erg from the initial fast rotator track of \cite{JohnstoneBartel2021} as an upper bound on the XUV flux. Repeating our calculations for both maximum mass loss rates and heating efficiencies, we find that TOI-1259~A~b has lost at most 0.61$\%$ of its initial mass to photoevaporation. These calculations assume that the planet either formed in situ, or migrated very early (i.e., before the dissipation of the gas disk). The unique architecture of this system instead indicates that this planet could have migrated in relatively late via the Kozai mechanism (see Sec.~\ref{sec:kozai}).  If so, its integrated total mass lost to date should be even lower.

This calculation tells us about the planet's past mass loss history. However, we can also look forward in time and calculate whether or not TOI-1259~A~b will survive over the remaining main sequence lifetime of it's host star, assuming a fixed mass loss rate of $10^{10.40}$~g~s$^{-1}$ (HD 40307) and $10^{10.65}$~g~s$^{-1}$ ($\epsilon$ Eridani) (we note that this is likely an overestimate as the stellar activity of the star should decrease with age, reducing $\dot{M}$). We find that it would take between 594-1,060~Gyr for photoevaporation to strip the planet of its envelope, and conclude that the planet should remain stable against photoevaporation over the star's remaining main sequence lifetime. Our estimate is comparable to the predicted atmospheric lifetimes of other gas-giant planets undergoing photoevaporation near the upper edge of the Neptune desert  \citep{VissapragadaKnutson2022}. 

Recent studies \citep[e.g.][]{LamponLopez-Puertas2023} have proposed that some escaping atmospheres can have super-solar hydrogen to helium ratios (e.g., 99/1 H/He), potentially resulting in mass loss rates that are an order of magnitude larger for the same helium absorption signal. We evaluate the effect of a 99/1 H/He atmospheric composition on our estimated mass loss rate for TOI-1259~A~b by running a Parker wind model with this alternative composition. We find a $\log{\dot{M}} = 11.17^{+0.43}_{-0.73}$~g~s$^{-1}$, which is an order of magnitude larger than our mass loss rate for a 90/10 H/He atmosphere, consistent with the predictions of \citet{LamponLopez-Puertas2023}. Even with this higher mass loss rate, we find that the planet lost $<10\%$ of its initial mass and it would take approximately 58 Gyr to completely strip the planet’s atmosphere. We conclude that even if the outflow is significantly enhanced in hydrogen relative to helium, the planet should still retain its atmosphere over the main sequence lifetime of the host star.

\subsection{Investigating Kozai-Lidov Cycles}\label{sec:kozai}

The presence of a white dwarf companion to TOI-1259~A suggests that the planet might have formed on a more distant orbit and then underwent HEM. This could provide some additional protection against atmospheric mass loss, as this type of migration is typically quite slow ($\sim$Gyr timescales). If the planet was located on a more distant orbit at early times when the star was most active, it would result in a lower time-integrated mass loss rate than scenarios where the planet was already located on a close-in orbit when the gas disk dissipated. In this section, we evaluate the plausibility of this scenario in more detail.

When the white dwarf ($0.561~M_{\odot}$, $a\sim$\SI{1648}{\au}) companion was on the main sequence, it would have been more massive (1.59~$M_{\odot}$) and much closer to TOI-1259~A \citep[$a\sim$\SI{900}{\au};][]{MartinEl-Badry2021}. Consequently, the proto-white dwarf could have potentially induced secular effects, such as Kozai-Lidov oscillations, on the planet's orbit \citep{Kozai1962, Lidov1962, MazehShaham1979}. We follow a similar methodology to \cite{NgoKnutson2015} to evaluate the plausibility of this scenario.

Because Kozai cycles are controlled by relatively weak tidal forces, they can be easily suppressed by pericenter precession ($\dot{\omega_{GR}}$). This precession can be induced by a variety of different factors, including general relativistic effects from the host star, or by the gravitational interactions with a nearby planet on an adjacent orbit
\citep{FabryckyTremaine2007}. Radial velocity observations of TOI-1259 A do not find any evidence for the presence of additional nearby planets in the system \citep{MartinEl-Badry2021}. We therefore compare the timescale of Kozai oscillations:

\begin{equation}\label{eq:kozai_timescale}
    \tau = \frac{2 P_{out}^2}{3 \pi P_{in}}\frac{m_1 + m_2 + m_3}{m_3}(1 - e_{out}^{2})^{2/3}
\end{equation}

to the timescale of pericenter precession due to general relativity:

\begin{equation}\label{eq:pericenter_timescale}
    \tau_{GR} = \frac{1}{\dot{\omega_{GR}}} = \frac{a_{in}^{5/2} c^2 (1-e_{in}^2)}{3 G^{3/2} (m_1 + m_2)^{3/2}}
\end{equation}

in which $P_{in}$ is the period of the inner binary (planet orbiting host star, approximately \SI{14}{\year} if the planet was originally at $\sim$5 AU), $P_{out}$ is the period of the outer binary (the planet-hosting star and its proto-white-dwarf companion; we adopt 6~$\times$~10$^4$~\si{\year}, approximated using Kepler's third law), $m_{1}$ is the mass of the host star \citep[0.68~M$_\odot$;][]{MartinEl-Badry2021}, $m_{2}$ is the mass of the planet \citep[0.441~M$_\mathrm{Jup}$;][]{MartinEl-Badry2021}, $m_{3}$ is the mass of the proto-white dwarf companion \citep[1.59~M$_\odot$;][]{MartinEl-Badry2021}, $e_{out}$ is the eccentricity of the outer binary (eccentricity of the proto-white dwarf), $a_{in}$ is the semi-major axis of the inner binary (the planet), $e_{in}$ is the eccentricity of the inner binary (assumed circular), $G$ is the gravitational constant and $c$ is the speed of light. We note that the eccentricity of the proto-white dwarf is unknown, and we therefore perform the above calculation for eccentricities varying between $0-1$. Additionally, we assume that the planet's initial orbital semi-major axis ($a_{in}$) was significantly farther out than its current value. Given that the orbital semi-major axis distribution of giant planets 
peaks at 2-\SI{6}{\au} \citep{FernandesMulders2019,FultonRosenthal2021,DrazkowskaBitsch2022}, we assume an initial $a_{in}$ of \SI{5}{\au} and calculate $P_{in}$ accordingly.

We next compare the two timescales. If the timescale of pericenter precession is shorter, Kozai cycles will be suppressed and the planet will remain at large orbital separations. We find that the timescale of pericenter precession always exceeds $\sim$~\SI{500}{\mega\year}, while the timescale of Kozai oscillations is always less than \SI{100}{\mega\year} regardless of the assumed orbital eccentricity for the proto-white dwarf. We therefore conclude that Kozai oscillations could have caused TOI-1259 A b to migrate inward from a more distant formation location. However, we note that other factors such as pericenter precession due to tides, stellar rotational distortions or perhaps the presence of other planetary bodies could also suppress Kozai oscillations \citep{FabryckyTremaine2007} and should be considered in a more detailed study.

\section{Conclusions}

In this work, we present the first detection of an escaping atmosphere on TOI-1259~A~b, a Saturn-mass planet orbiting a K dwarf with a white dwarf companion. Using Palomar/WIRC we measure an excess metastable helium absorption signal in the transit light curve of $0.395\pm{0.072}\%$. Using Keck/NIRSPEC we measure a blue-shifted peak excess absorption of $2.4\pm0.52\%$  between first and fourth contact and $3.5\pm0.72\%$  between second and third contact. Integrating our Keck/NIRSPEC signals with the Palomar/WIRC bandpass, we 
find that our second and third contact signal, which is more reflective of the behavior at mid-transit, is consistent with the Palomar/WIRC mid-transit excess absorption at the 1$\sigma$ level. We convert the Palomar/WIRC and Keck/NIRSPEC excess absorption measurements into constraints on the mass loss rate of the outflow  
and find that the planet is losing mass at a rate of $\log{\dot{M}} = 10.40^{+0.15}_{-0.20}$~g~s$^{-1}$ with our Palomar/WIRC measurement, and $\log{\dot{M}} = 10.2^{+0.10}_{-0.25}$~g~s$^{-1}$ ($T_1-T_4$) and $\log{\dot{M}} = 10.3^{+0.15}_{-0.25}$~g~s$^{-1}$ ($T_2-T_3$) with our Keck/NIRSPEC measurements. However, a single epoch of HIRES observations reveals a potentially active host star, so we repeat our mass loss calculations with a more active proxy stellar spectrum and find that at most, the planet is losing mass at a rate of ${\dot{M}} = 10^{10.65}$~g~s$^{-1}$. Using our maximum mass loss rate, we derive a corresponding predicted atmospheric lifetime ($M_{env}/~\dot{M}$) of $\sim$\SI{1000}{\giga\year} which is substantially longer than the lifetime of an average main sequence star. We also calculate the fraction of the initial planet mass lost to photoevaporation, assuming a constant mass loss efficiency, and find that the planet lost at most, only 0.61\% of its initial mass. Therefore the planet does not appear to be significantly sculpted by photoevaporation. This is consistent with results for other planets along the upper edge of the Neptune desert, which are also stable against photoevaporation and display similar mass loss rates and heating efficiencies \citep{VissapragadaKnutson2022, DosSantos2022_observations}. 

Given the presence of the white dwarf companion, this planet is also a potential candidate for HEM via Kozai-Lidov oscillations induced by the proto-white dwarf. We compare the timescale of Kozai oscillations to the timescale of pericenter precession and find that the timescale of Kozai oscillations is always less than that of pericenter precession. We conclude that this planet could have formed on a more distant orbit and then migrated inward, which would further protect it from atmospheric mass loss. Future studies of this system could additionally investigate this scenario by obtaining improved constraints on the planet's orbital eccentricity. Although tidal effects are expected to circularize its orbit over time, the presence of a remnant eccentricity would strengthen the evidence for HEM. 
 
If TOI-1259~A~b formed in a more distant formation location, evidence for this should also be recorded within TOI-1259~A~b's atmosphere. Because HEM can only operate after the gas disk has dissipated, we can expect that TOI-1259 A b’s atmosphere should have been minimally affected by migration-driven accretion and should therefore preserve a pristine record of its formation location. As discussed in \citet{ObergMurray-Clay2011}, \citet{madhusudhan_co_2012}, and \citet{ChachanKnutson2023} the abundance of refractory elements (e.g. the carbon-to-oxygen ratio C/O) in the planet's atmosphere can be used as a tracer of formation location.
If TOI-1259 A b formed outside of
the water snowline, its atmosphere would have been enriched in oxygen and enhanced by the accretion
of small water-rich solids, therefore resulting in a substellar C/O ratio. If TOI-1259~A~b instead formed interior to the water ice line, it would have accreted
more carbon-rich solids and other refractory elements such as sulfur, which would be reflected in an enhanced refractory element abundance \citep{turrini_tracing_2021, lothringer_new_2021, pacetti_chemical_2022, Crossfield2023, TsaiLee2023}. 
TOI-1259~A~b has an equilibrium temperature of 917 K, and we therefore expect that its atmosphere may contain detectable quantities of H$_2$O, CO, CO$_2$, SO$_2$ and CH$_4$. 
If we can measure the abundances of most or all of these species in its atmosphere using transmission spectroscopy, we can potentially determine where in the disk TOI-1259~A~b formed. TOI-1259~A~b has a Transmission Spectroscopy Metric \citep[TSM;][]{KemptonBean2018} of 184, one of the highest TSMs for a Jupiter-sized
planet cooler than 1000~K, rendering TOI-1259~A~b a compelling target for future atmospheric characterization studies.

\section*{acknowledgments}

We acknowledge the use of public TESS data from pipelines at the TESS Science Office and at the TESS Science Processing Operations Center. This paper includes data collected by the TESS mission that are publicly available from the Mikulski Archive for Space Telescopes (MAST). Funding for the TESS mission is provided by NASA's Science Mission Directorate. The specific observations analyzed can be accessed via \dataset[https:/doi.org/10.17909//t9-nmc8-f686]{https:/doi.org/10.17909/t9-nmc8-f686} \citep{MASTdoi}. STScI is operated by the Association of Universities for Research in Astronomy, Inc., under NASA contract NAS5–26555. Support to MAST for these data is provided by the NASA Office of Space Science via grant NAG5–7584 and by other grants and contracts. This research has made use of the NASA Exoplanet Archive, which is operated by the California Institute of Technology, under contract with the National Aeronautics and Space Administration under the Exoplanet Exploration Program. This material is based upon work supported by the National Science Foundation Graduate Research Fellowship Program under Grant No.~DGE‐1745301. Any opinions, findings, and conclusions or recommendations expressed in this material are those of the author(s) and do not necessarily reflect the views of the National Science Foundation. 

We thank the Palomar Observatory telescope and support operators for their support of this work, with special thanks to Tom Barlow, Carolyn Heffner, Isaac Wilson, Diana Roderick, and Joel Pearman. We also thank the Keck Observatory telescope and support astronomers, with special thanks to Joel Aycock and Percy Gomez.

\vspace{5mm}
\facilities{ADS, NASA Exoplanet Archive, Hale 200-inch (WIRC), Keck II/NIRSPEC}

\software{\texttt{astropy} \citep{AstropyCollaborationRobitaille2013, AstropyCollaborationPrice-Whelan2018, AstropyCollaborationPrice-Whelan2022}, 
          \texttt{numpy} \citep{HarrisMillman2020}, \texttt{scipy} \citep{VirtanenGommers2020}, \texttt{matplotlib} \citep{Hunter2007}, \texttt{exoplanet} \citep{Foreman-MackeyLuger2021, Foreman-MackeyLuger2021zenodo}, \texttt{starry} \citep{LugerAgol2019}, \texttt{lightkurve} \citep{LightkurveCollaborationCardoso2018}, \texttt{pymc3} \citep{SalvatierWieckia2016}, \texttt{theano} \citep{TheTheanoDevelopmentTeamAl-Rfou2016}, \texttt{p-winds}\citep{DosSantosVidotto2022}, \texttt{Cloudy} \citep{FerlandKorista1998, FerlandChatzikos2017}, \texttt{lmfit} \citep{NewvilleStensitzki2016}, \texttt{sunbather}\citep{LinssenOklopcic2022}
          }

\bibliography{references}{}
\bibliographystyle{aasjournal}

\end{document}